\RequirePackage{fix-cm}
\documentclass[smallextended]{svjour3}       
\smartqed  
\usepackage{graphicx}
\usepackage{natbib}
\usepackage{fancyhdr}
\usepackage{float}
\usepackage{amssymb}
\usepackage{latexsym}
\usepackage{amsmath, bm,bbm}
\usepackage{amsfonts}
\usepackage{rotating}
\usepackage{caption}
\usepackage{color,soul}
\usepackage{hyperref}
\hypersetup{colorlinks=true, linkcolor=blue, urlcolor=blue, citecolor=blue}
\usepackage{enumerate}
\usepackage{epsfig}
\usepackage{subfigure}
\usepackage{setspace}
\usepackage{appendix}
\usepackage{multirow}
\usepackage{booktabs,caption,fixltx2e}
\usepackage[flushleft]{threeparttable}
\usepackage{dsfont}
\usepackage{url}  
\usepackage[margin=1in]{geometry}
\usepackage{authblk}
\usepackage{pdflscape}
\begin{document}

\title{Estimation of causal effects of multiple treatments in healthcare database studies with rare outcomes\thanks{Corresponding author: Liangyuan Hu, Department of Population Health Science and Policy, Icahn School of Medicine at Mount Sinai, New York, NY, 10029, USA. Email: liangyuan.hu@mssm.edu. }
}


\author{Liangyuan Hu        \and
        Chenyang Gu 
}


\institute{Liangyuan Hu \at
             Department of Population Health Science and Policy, Icahn School of Medicine at Mount Sinai, New York, NY, 10029, USA \\
              \email{liangyuan.hu@mssm.edu}           
           \and
           Chenyang Gu \at
              Analysis Group, Inc., Los Angeles, CA, 90071, USA\\
               \email{Chenyang.Gu@analysisgroup.com} 
}

\date{Received: date / Accepted: date}

\maketitle

\begin{abstract}
The preponderance of large-scale healthcare databases provide abundant opportunities for comparative effectiveness research. Evidence necessary to making informed treatment decisions often relies on comparing effectiveness of multiple treatment options on outcomes of interest observed in a small number of individuals. Causal inference with multiple treatments and rare outcomes is a subject that has been treated sparingly in the literature. This paper designs three sets of simulations, representative of the structure of our healthcare database study, and propose causal analysis strategies for such settings.
We investigate and compare the operating characteristics of  three types of methods and their variants:  Bayesian Additive Regression Trees (BART), regression adjustment on multivariate spline of generalized propensity scores (RAMS) and inverse probability of treatment weighting (IPTW) with multinomial logistic regression or generalized boosted models.  Our results suggest that BART and RAMS  provide lower bias and mean squared error, and the widely used IPTW methods deliver unfavorable operating characteristics. We illustrate the methods using a case study evaluating the comparative effectiveness of robotic-assisted surgery, video-assisted thoracoscopic surgery and open thoracotomy for treating non-small cell lung cancer.
\end{abstract}

\section{Introduction}

\subsection{A comparative effectiveness study}
Robotic-assisted surgery (RAS), a minimally invasive technology that uses robotic system to enhance instrument maneuverability and visualization, was introduced in 2002 to treat resectable non-small cell lung cancer (NSCLC)  \citep{melfi2002early}. This technology enables surgeons to perform intricate maneuvers with greater precision \citep{veluswamy2019comparative}. There are no randomized controlled trials (RCTs) that have been conducted to compare the effectiveness of RAS with commonly used surgical approaches. Current comparative effectiveness studies have been relying on healthcare databases.

Several studies reported shorter length of stay, reduced pain, less postoperative bleeding and lower rate of myocardial infarction among NSCLC patients who were treated with RAS compared to those who were operated via open thoracotomy (OT) \citep{veronesi2010four, cerfolio2011initial,oh2017robotic},  and similar survival compared to those who received video-assisted thoracoscopic surgery (VATS), another minimally invasive technology \citep{farivar2014comparing}. With more evidence supporting its feasibility and safety, RAS has been rapidly adopted since 2008 \citep{novellis2017robotic}. The lack of randomization however contributed to diverging results. For example, \cite{veluswamy2019comparative} showed that RAS led to similar post-operative cardiovascular complication rate as VATS and OT, and recommended that comparative effectiveness of RAS should be further evaluated prior to widespread adoption. 

Previous comparative research of theses surgical approaches is limited with respect to the statistical analysis approach. Existing studies primarily focused on pairwise comparison between two surgical approaches for patients who are eligible for the specific pair of surgical procedures \citep{cajipe2012video, farivar2014comparing}. Findings from these studies, in general, cannot inform the optimal surgical method for the entire population who are eligible for all three surgical procedures \citep{lopez2017estimation}. \cite{veluswamy2019comparative} used a matched study cohort that was much smaller than the sample population, limiting the generalizability of their study results. 

\subsection{Research gaps and objectives}
The healthcare databases for this comparative question pose additional statistical challenges. One important outcome for assessing the effectiveness of different surgical management methods is postoperative morbidity (e.g., extrapulmonary infection, cardiovascular complication, etc.). However, only a small fraction of patients may have these complications after surgery \citep{veluswamy2019comparative}.  For example, less than 3\% of NSCLC patients who underwent surgeries experienced cardiovascular complication, and less than 2\% had reoperation. The small number of outcome events in these surgical groups implies limited information on the comparison of the surgical approaches.  As such, we need to compare a large amount of patients receiving different treatment regimens with respect to the risk of the health event of interest, while adjusting for many potential confounders. This is particularly challenging with rare outcomes because the great majority of patients contribute no information to explaining the variability attributable to the differential treatment regimens in the health events. Other examples of rare outcomes include studies of rare disease or rare adverse safety event or new treatment. Despite the challenge on analytic efficiency posed by rare outcomes, research on the performance of causal inference techniques in this context is sparse with the exception of  \cite{cepeda2003comparison} and  \cite{franklin2017comparing}, which applied and compared different propensity score methods in the context of rare outcomes and a binary treatment. 
 \citet{cepeda2003comparison} conducted simulation studies to compare logistic regression with propensity score methods when the number of events is low relative to the number of confounders, and found that the propensity score methods had more empirical power than logistic regression. \citet{franklin2017comparing} further described and compared a multitude of propensity score-based approaches in healthcare database studies with rare outcomes in the context of binary treatment. Their simulation study concluded that regression adjustment on propensity score using one nonlinear spline (RAONS) performed best with respect to bias and RMSE in the absence of treatment heterogeneity.

Causal inference techniques have traditionally focused on binary treatment. There has been growing literature developing regression and propensity score-based  techniques with multiple treatments in recent years. These methods were developed with continuous outcomes in mind and were less studied for binary outcomes. \cite{lopez2017estimation} provided a comprehensive review of
current methods for multiple treatments. \cite{hu2020estimation} explicated use of methods appropriate for multiple treatments in the context of binary outcomes (with prevalence rate between 10\% to 75\%) and presented empirical evidence about the performances of several propensity score- and machine learning-based methods in a variety of complex causal inference settings, considering sample size, ratio of units, non-linearity and non-additivity
of both the treatment assignment and outcome generating mechanisms and levels of covariate overlap. Sets of simulations in \cite{hu2020estimation} suggest that Bayesian Additive Regression Trees (BART) -- a nonparametric modeling technique -- provides best performance on the bases of better bias reduction, smaller root mean squared error (RMSE), more consistent 95\% uncertainty coverage and better large-sample convergence property, compared to inverse probability of treatment weighting (IPTW)\citep{feng2012generalized}, regression adjustment (RA) \citep{linden2016estimating}, generalized boosted models (GBM) \citep{mccaffrey2013tutorial}, vector matching \citep{lopez2017estimation}, targeted maximum likelihood estimation (TMLE), \citep{rose2019double} and IPTW with super learner \citep{van2007super}.

To our knowledge, there is no literature comparing causal inference approaches in the context of both rare outcomes and multiple treatments.To fill in research gaps in causal inference literature in the setting of multiple treatments and rare outcomes, we first select the top performer in each component of the setting, i.e.,BART for the multiple treatment setting and RAONS for the binary treatment and rare outcomes setting, and then extend the two methods in the context of multiple treatments and binary outcomes.  We propose a new strategy, regression adjustment on multivariate spline of the generalized propensity score (GPS) (RAMS), which extends RAONS to accommodate multiple treatments by using a nonlinear spline model for the outcome that is additive in the treatment and multivariate spline function of the GPS. In addition, we include IPTW as a comparison method because it is widely used in the causal analysis of multiple treatments.  We further generate simulation scenarios that are uniquely representative of studies constructed from large scale healthcare databases and provide empirical evidence regarding the comparison of the three types of methods and their varied versions via extensive simulations. Our results suggest 
BART consistently provides the best performance corresponding to bias and RMSE across all scenarios considered, RAMS has similar performance under certain settings with the advantage of computational efficiency, and the widely used IPTW methods deliver poor operating characteristics in all conditions. 

This paper is organized as follows: the remainder of Section 1 provides additional background on the healthcare database study.  Section 2 introduces the potential outcomes framework for multiple treatments and the methods for estimating causal effects with multiple treatments. Section 3 presents the design and results of the Monte Carlo simulation. Section 4 describes the empirical data analysis. Conclusions and discussions are provided in Section 5.

\subsection{Data source: SEER-Medicare linkage}
To demonstrate the methods, we conduct a case study constructed from the latest SEER-Medicare linkage to compare the effects of RAS, VATS and OT among early stage NSCLC patients. The  data set includes a total of 11980 patients above 65 years of age with stage I-IIIA NSCLC, diagnosed between 2008 (first year patients in the registry underwent robotic-assisted surgery) and 2013, and underwent one of the three surgical procedures.  Among these patients, 396 (3.3$\%$), 6582 (54.9$\%$) and 5002 (41.8$\%$) received RAS, VATS and OT, respectively. The data set contains patient-level pre-treatment information on socio-demographics (including age, gender, marital status, race, income level), year of diagnosis, comorbidities, tumor characteristics and stage (including cancer stage, tumor size, tumor site, histology), and  preoperative evaluation (including positron emission tomography [PET] scan, chest computer tomography or mediastinoscopy). We use the postoperative morbidity as the outcome, which is defined as the presence of complications within 30 days of surgery or during the hospitalization in which surgery was performed \citep{veluswamy2019comparative}. In this study, four types of complications that are rare from Medicare claims are used: 1) extrapulmonary infections; 2) cardiovascular complications; 3) thromboembolic events; 4) reoperations. Table \ref{Tab.Seer} summarizes baseline characteristics and postoperative complications for the overall population and each surgical group.

\begin{table}[H]
\centering
\scriptsize
\caption{\small Baseline characteristics of subjects receiving three surgical procedures in the SEER-Medicare dataset. SD: standard deviation. OT:  Open Thoracotomy. RAS:  Robotic-Assisted Surgery. For statistical testing,  regular ANOVA was used for continous variables. Chi-squared test was used for categorical variables. $\dagger$: the number of events under 11 was masked according to the CMS cell size suppression policy.}\label{Tab.Seer}

\begin{tabular}{lccccc}
\hline
&  Overall  & VATS & OT & RAS& $p$-value \\ 
Variables & $N = 11980$  & $N = 6582$ & $N = 5002$  &  $N = 396$ \\
\hline
\\
\textbf{\textit{Preoperative characteristics complications}} &  &  &  & \\
Age (years), mean (SD) & 74.2 (5.6)  & 73.9 (5.4) & 74.5 (5.7) &  74.3 (5.7) & 0.077\\ 
Female, N ($\%$) &  6610 (55.2)  & 3446 (52.4) & 2941 (58.8) &  223 (56.3)& 0.011\\ 
Married, N ($\%$) & 6782 (56.6)   & 3753 (57.0) & 2802 (56.0) & 227 (57.3) & 0.537\\ 
Race, N ($\%$) &  &  &  &  & \\   
\;\; White & 10383 (86.7)  & 5694 (86.5) & 4369 (87.3) & 320 (80.8) & 0.056 \\  
\;\; Black & 633 (5.3)  & 364 (5.5) & 248 (5.0) & 21 (5.3)& 0.395\\   
\;\;  Hispanic & 372 (3.1) & 218 (3.3) & 139 (2.8) &  15 (3.8) & 0.190 \\   
\;\;  Other &  592 (4.9) & 306 (4.6) & 246 (4.9) & 40 (10.1) & $<$0.001 \\ 
Median household annual income, N ($\%$) &  &  &  &  & \\   
\;\; 1st quartile &3238 (27.0)  & 2132 (32.4) & 1009 (20.2) & 97 (24.5)&$<$0.001\\   
\;\; 2nd quartile &  3010 (25.1)& 1729 (26.3) & 1193 (23.9) & 88 (22.2)& 0.042\\   
\;\; 3rd quartile &  2586 (21.6)& 1345 (20.4) & 1143 (22.9) & 98 (24.7)& 0.022 \\   
\;\; 4th quartile &  3146 (26.3) & 1376 (20.9) & 1657 (33.1) &113 (28.5)& $<$0.001\\ 
Charlson comorbidity score, N ($\%$) &  &  &  &  & \\  
\;\;  $0-1$ & 4127 (34.4)  & 2163 (32.9) & 1810 (36.2) & 154 (38.9) & 0.005\\ 
\;\;   $1-2$ &  3436 (28.7) & 1944 (29.5) & 1379 (27.6) &113 (28.5) & 0.068\\  
\;\;   $>2$ &  4417 (36.9) & 2475 (37.6) & 1813 (36.2) & 129 (32.6)& 0.064\\ 
Year of diagnosis, N ($\%$) &  &  &  &  & \\  
\;\;  2008-2011 & 8181 (68.3)  & 4842 (73.6) & 3207 (64.1) & 132 (33.3) & $<$0.001\\  
\;\;  2012 &  1851 (15.5) & 899 (13.7) & 821 (16.4) & 131 (33.1) & $<$0.001\\   
\;\;  2013 &  1948 (16.3) & 841 (12.8) & 974 (19.5) & 133 (33.6)& $<$0.001\\ 
Cancer stage, N ($\%$) &  &  &  &  & \\  
\;\;   Stage I & 8374 (69.9) & 4195 (63.7) & 3884 (77.6) &  295 (74.5) & $<$0.001\\   
\;\;  Stage II & 2276 (19.0) & 1504 (22.9) & 709 (14.2) &  63 (15.9) & $<$0.001\\ 
\;\;  Stage IIIA & 1330 (11.1) & 883 (13.4) & 409 (8.2) &  38 (9.6) & $<$0.001\\ 
Tumor size, in mm, N ($\%$) &  &  &  &  & \\  
\;\;  $\leq 20$ & 4359 (36.4)  & 1967 (29.9) & 2232 (44.6) & 160 (40.4) & $<$0.001\\ 
\;\;  $21-30$ & 3182 (26.6)  & 1696 (25.8) & 1388 (27.7) & 98 (24.7) & 0.041\\ 
\;\;  $31-50$ & 2900 (24.2)  & 1804 (27.4) & 987 (19.7) & 109 (27.5) & $<$0.001\\ 
\;\;   $\geq 51$ & 1539 (12.8)  & 1115 (16.9) & 395 (7.9) & 29 (7.3) & $<$0.001\\ 
Histology, N ($\%$) &  &  &  &  & \\  
\;\;  Adenocarcinoma &7360 (61.4)  & 3757 (57.1) & 3348 (66.9) & 255 (64.4) &  $<$0.001\\   
\;\; Squamous cell carcinoma & 3439 (28.7)  & 2165 (32.9) & 1167 (23.3) & 107 (27.0) &  $<$0.001\\ 
\;\;  Other histology & 1181 (9.9)  & 660 (10.0) & 487 (9.7) &34 (8.6)  & 0.601\\
Tumor site, N ($\%$) &  &  &  &  &  \\  
\;\;  Upper lobe &  6903 (57.6)  & 3829 (58.2) & 2859 (57.2) &215 (54.3) & 0.216 \\  
\;\; Middle lobe & 670 (5.6)  & 308 (4.7) & 335 (6.7) & 27 (6.8) & 0.103\\    
\;\; Lower lobe & 4056 (33.9)  & 2195 (33.3) & 1720 (34.4) & 141 (35.6) & 0.381\\ 
\;\; Other site & 351 (2.9)  & 250 (3.8) & 88 (1.8) & 13 (3.3) & 0.078\\ 
PET scan, N ($\%$) & 8716 (72.8) & 5004 (76.0) & 3410 (68.2) &  302 (76.3) & $<$0.001\\ 
Chest computer tomography, N ($\%$) & 7936 (66.2)  & 4525 (68.7) & 3148 (62.9) &  263 (66.4) & $<$0.001\\ 
Mediastinoscopy, N ($\%$) &  1197 (10.0) & 715 (10.9) & 420 (8.4) &  62 (15.7) & $<$0.001\\
\\
 \textbf{\textit{Postoperative complications}} &  &  &  & \\
 Extrapulmonary infection& 614 (5.1)  & 406 (3.9)  & 195 (6.2)& 13 (3.3) & $<$0.001   \\
Cardiovascular complication$\dagger$ & 270 (2.3) & 160 (2.0)& $>90$ ($>2$)& $<11$ ($<3$) & 0.281  \\
Thromboembolic complication & 559 (4.7) & 302 (4.7) &236 (4.6)& 21 (5.3) & 0.786   \\
Reoperation$\dagger$ & 144 (1.2) & 100 (0.8) &$<50$ ($>1$)& $<11$ ($<2$) & 0.043  \\
 \hline
\end{tabular}
\end{table}

\section{Methods}

\subsection{Potential outcomes framework for multiple treatments}
Our notation is based on the potential outcomes framework, which was originally proposed by \citet{Neyman1990application} in randomized experiments with randomization-based inference, and generalized to observational studies and Bayesian analysis by \citet{rubin1974estimating, rubin1977assignment, rubin1978bayesian}, commonly referred to as the Rubin Causal Model \citep{holland1986statistics}. 

Suppose we are interested in evaluating the causal effect of a treatment or exposure $W$ on outcome $Y$ for a sample of units, indexed by $i = 1,\ldots,N$, drawn randomly from a target population. Each unit is exposed to one of $Z$ possible treatments; that is, $W_i = w$ if individual $i$ was observed under treatment $w$, where $w \in \mathcal{W} = \{1,2,\ldots,Z\}$. The number of units receiving treatment $w$ is $N_{w}$ and $\sum_1^Z N_{w} = N$. Let $Y_i$ be the observed outcome of the $i$th unit given the assigned treatment, and $\{Y_i(1),\ldots,Y_i(Z)\}$ the potential outcomes for the $i$th unit under each possible treatment of $\mathcal{W}$. Here we assume the stable unit treatment value assumption (SUTVA) \citep{rubin1980randomization}, that is, no interference between units and no different versions of a treatment. The observable outcome, $Y_i$, for unit $i$ can be written as 
\begin{equation} \label{eq: obs-po-y}
Y_i = \sum_{w \in\{1,2,\ldots,Z\}} Y_i(w) I(W_i = w),
\end{equation}
where $I(\cdot)$ is an indicator function. Known as the fundamental problem of causal inference, for each unit, at most one of the potential outcomes is observed (the one corresponding to the treatment to which the unit is exposed), and all other potential outcomes are missing \citep{holland1986statistics}.

Causal effects are summarized by estimands, which are functions of the unit-level potential outcomes on a common of set of units \citep{rubin1974estimating, rubin1978bayesian}. Here we focus on a common estimand in comparative effectiveness studies of binary outcomes, marginal risk difference(RD), in the multiple treatment setting. There are two types of estimands, average treatment effect (ATE) and average treatment effect on the treated (ATT).  We focus on ATE in this paper as it is of great interest to generalize inference about the effectiveness of the surgical approaches to the overall population. Implementation of all methods considered for ATT estimation is straighforward. The ATE in terms of marginal RD in the sample for treatment $k,l \in \mathcal{W}$ is defined as 
\begin{equation}
\Delta_{k,l} = \sum_{i=1}^N Y_i(k) - \sum_{i=1}^N Y_i(l).
\end{equation}

In general, causal effects are not identifiable without further assumptions because only one of the potential outcomes is observed for each unit. The key assumptions concern the assignment mechanism, or in other words, how each unit received the treatment it actually received \citep{imbens2015causal}. The first assumption is known as positivity or common support (overlap) assumption. More specifically, 
\begin{equation*}
0 < p(W_i|Y_i(1),\ldots,Y_i(Z),\bm{X}_i) < 1, \quad \forall \; w \in \{1,\ldots,Z\},
\end{equation*}
where $\bm{X}_i$ is a vector of pre-treatment covariates. We further assume that, given the pre-treatment covariates, $\bm{X}_i$, the assignment mechanism does not depend on the potential outcomes, so that it is unconfounded \citep{imbens2015causal}:  
\begin{equation}
\label{gps}
p(W_i|Y_i(1),\ldots,Y_i(Z),\bm{X}_i) = p(W_i|\bm{X}_i), \quad \forall \; w \in \{1,\ldots,Z\}.
\end{equation}
This assumption implies that the set of observed pre-treatment covariates, $\bm{X}_i$, is sufficiently rich such that it includes all variables directly influencing both $W_i$ and $Y_i$; in other words, there is no unmeasured confounding. The probability in equation~\ref{gps} is referred to as the GPS. For each unit, there is a vector of propensity scores corresponding to each treatment, $\bm{R}(\bm{X}_i) = (r(1,\bm{X}_i),\ldots, r(Z,\bm{X}_i))$, where $r(w,\bm{X}_i) = p(W_i=w | \bm{X}_i)$, for $w \in \mathcal{W}$.

We also define the response surface as follows,
\begin{equation}
f(w,\bm{X}_i) \equiv E[Y_i(w)|\bm{X}_i],  \quad \forall \; w \in \{1,\ldots,Z\}.
\end{equation}
Under the assumption of unconfoundedness and common support, the response surface
for any $w$ can be evaluated by the mean function of the observed outcomes conditional on treatment and pre-treatment covariates $\bm{X}_i$, 
\begin{equation}
f(w,\bm{X}_i) = E[Y_i(w) | W=w, \bm{X}_i] = E[Y_i | W=w, \bm{X}_i],
\end{equation}

\subsection{Overview of methods considered for rare outcomes and multiple treatments} \label{sec:approaches}

\subsubsection{Inverse probability of treatment weighting}
A widely used method to estimate causal effects with multiple treatments is the inverse probability of treatment weighting (IPTW) \citep{rosenbaum1987model, robins2000marginal, hu2019causal}. The concept of IPTW was originally introduced by \citet{horvitz1952generalization} in survey research to adjust for imbalances in sampling pools. Weighting methods have been extended to estimate causal effect of a binary treatment in observational studies, and recently reformulated to accommodate multiple treatments \citep{imbens2000role, feng2012generalized, mccaffrey2013tutorial}. The weighting estimator of $\Delta_{k,l}$ is written as follows:
\begin{equation}
\widehat{\Delta}^{iptw}_{k,l} =  \frac{\sum_{i=1}^NY_i I(W_i = k) / \hat{r}(k,\bm{X}_i)}{\sum_{i=1}^N Y_i I(W_i = l) /\hat{r}(l,\bm{X}_i)},
\end{equation}
where $\hat{r}(w,\mathbf{X}_i)$ is the estimated GPS, for $w \in \mathcal{W}$. In practice, the GPS is often estimated from multinomial logistic regression. A challenge of the IPTW method is that treated units with the estimated GPS that are close to zero would result in  extreme weights, which lead to unstable causal estimates with large sample variances \citep{little1988missing, kang2007demystifying}. This issue is more likely as the number of treatments increases, as treatment assignment probabilities for some treatment groups may become quite small \citep{lopez2017estimation}. One remedy is weight trimming, or weight truncation, \citep{cole2008constructing, lee2011weight}, by which extreme weights that fall outside a specified range limit of the weight distribution are set to the range limit. The limit is often based on percentiles of the weight distribution (e.g., the $5^{th}$ and $95^{th}$ percentiles).

An alternative method to estimating the GPS is the generalized boosted models (GBM) \citep{friedman2002stochastic,mccaffrey2013tutorial}. GBM is a machine learning technique and has been utilized in the binary treatment setting to flexibly estimate complex and nonlinear relationship between treatment assignment and pre-treatment covariates without overfitting the data \citep{mccaffrey2004propensity}. GBM has an iterative estimation procedure that can be tuned to find the propensity score model  producing the best pre-treatment covariate balance between treatment groups. Previous studies show that the use of machine learning techniques such as GBM in propensity score estimation is able to alleviate extreme weights and improve the estimation of causal effects \citep{lee2010improving, mccaffrey2013tutorial}. However, the iterative algorithm of GBM can be computationally intensive when the sample size or the number of pre-treatment covariates is large.

\subsubsection{Bayesian Additive Regression Trees}
Another causal inference approach is to directly model  the response surface $f(w,\bm{X})$. This method is also known as regression adjustment (RA) \citep{rubin1973use, rubin1979using} or model-based imputation \citep{imbens2015causal}. 
Potential outcomes corresponding to treatment $k$ and $l$ for each unit are imputed, respectively, from the regression model of $f(w,\bm{X})$, and then averaged and contrasted to estimate the ATE effect $\Delta_{k,l}$.  The critical part of this method is the specification of the functional form of the  regression model. Model misspecification  could lead to biased treatment effect estimates. 

Given advances in Bayesian nonparametric models with highly flexible functional
form,  BART, has been used to model the response surface $f(w,\bm{X})$ \citep{chipman2007bayesian, chipman2010bart, hill2011bayesian, hu2020estimation}. 
 BART is a Bayesian ensemble method that models the conditional mean outcome  by a sum of regression trees. \citet{chipman2010bart} has provided extensive evidence to demonstrate that BART has excellent predictive performance and is well-suited to detecting nonlinearities and interactions with little parameter tuning and coherent uncertainty estimation. \citet{hill2011bayesian} proposed the use of BART for causal inference with a binary treatment. \citet{hu2020estimation} extended BART to estimate the causal effects of multiple treatments in observational studies with a binary outcome. For a binary outcome, the BART model can be formulated to the probit model setup
\begin{equation}
f(w,\bm{X}_i) = E[Y_i | W_i = w, \bm{X}_i] = \Phi \bigg{\{} \sum_{j=1}^J g_j(w, \bm{X}_i; T_j, M_j) \bigg{\}},
\end{equation}
where $\Phi$ is the the standard normal c.d.f., each $(T_j, M_j)$ denotes a single subtree model in which $T_j$ denotes the regression tree and $M_j$ is a set of parameter values associated with the terminal nodes of the $j$th regression tree, $g_j(w,\bm{X}_i; T_j, M_j)$ represents the mean assigned to the node in the $j$th regression tree associated with covariate value $\bm{X}_i$ and treatment level $w$, and the number of regression trees $J$ is considered to be fixed and known. The details of prior specification and Bayesian backfitting algorithm for posterior sampling can be found in \citet{chipman2010bart}. A total of $S$ Markov Chain Monte Carlo samples of model parameters, $(T_j, M_j)$, are drawn from their posterior distribution. The missing potential outcomes are predicted for each unit and the relevant treatment level in each of $S$ draw. The average treatment effect $\Delta_{k,l}$ comparing treatment group $k$ and $l$ can be estimated as follows:
\begin{equation}
\widehat{\Delta}^{bart}_{k, l} =  \frac{\sum_{s=1}^S \sum_{i=1}^N    f^s(k, \bm{X}_i)}{\sum_{s=1}^S \sum_{i=1}^N f^s(l, \bm{X}_i)}  = \frac{\sum_{s=1}^S \sum_{i=1}^N  \Phi \bigg{\{} \sum_{j=1}^J g_j(k, \bm{X}_i; T^s_j, M^s_j) \bigg{\}}}{\sum_{s=1}^S \sum_{i=1}^N \Phi \bigg{\{} \sum_{j=1}^J g_j(l, \bm{X}_i; T^s_j, M^s_j) \bigg{\}}},
\end{equation} 
where $(T^s_j,M^s_j)$ are the $s$th draw from the posterior distribution of $(T_j,M_j)$. The point and interval estimates of average treatment effects can be obtained using the summary of posterior samples.

Unlike the propensity score-based methods, BART is not equipped with a mechanism that prevents it from extrapolating over the areas of the covariate space where common support does not exist. \cite{hu2020estimation} proposed a strategy to identify a common support region for retaining inferential units, and provided empirical evidence that the BART-specific discarding rule largely improved over BART based on bias reduction and RMSE by excluding units that fall outside the region of common support when there is lack of covariate overlap. Their strategy discards units that have large variability in predicting potential outcomes. Specifically, any unit $i$ with $W_i = w$ will be discarded if 
\begin{equation}
\label{eq:disc}
s_i^{f_{w'}} > \text{max}_j \{s_j^{f_w} \}, \forall j: W_j = w,  \; \text{and } w' \neq w \in \mathcal{W},
\end{equation}
where $s_j^{f_w}$ and $s_j^{f_w'}$ denote the standard deviation of the posterior distribution of the potential outcomes under treatment $W = w$ and $W=w'$, respectively, for a given unit $j$. When estimating the ATE, the discarding rule in \eqref{eq:disc} is applied to each treatment group. 

\subsubsection{Regression Adjustment with Multivariate Spline of GPS}
Regression adjustment with the PS is an alternative to the commonly used PS-based methods using weighting, matching, and stratification \citep{hade2014bias, vansteelandt2014regression}. This method involves a regression model for the outcome with the treatment and PS being the independent variables. However, the treatment effects estimated using this method may be biased when the regression model is misspecified, even if the PS model is correct \citep{cangul2009testing, vansteelandt2014regression}. One way to overcome  this problem is to allow for a flexible regression model to capture the nonlinear association between the outcome and the PS via a spline function (e.g., penalized spline of the PS) \citep{gutman2013robust, hade2014bias, vansteelandt2014regression}. \citet{franklin2017comparing} used expansive simulations to show that this method performed best corresponding to bias and RMSE compared to a multitude of PS-based methods  in the context of rare outcome.

Building on the work by \cite{franklin2017comparing}, we extend this method into the multiple treatment setting using a multivariate spline of the GPS. More specifically, with a binary outcome, we assume a nonlinear relationship between the outcome and the multivariate spline of the GPS as follows:
\begin{equation}
f(w,\bm{X}_i) =  E[Y_i | W_i = w, \bm{X}_i]  =  \text{logit}^{-1} \bigg{\{}  \bm{\beta} W_i + h(\bm{R}(\bm{X}_i),\bm{\gamma}) \bigg{\}},
\end{equation}
where $h(\cdot)$ is some spline function of the GPS indexed by $\bm{\gamma}$ and $\bm{\beta} = [\beta_1, \ldots, \beta_Z]^T$ are regression coefficients associated with the treatment $W_i$. The dimension of the spline function $h(\cdot)$ depends on the number of treatments $Z$. For example, when $Z=3$, $h(\cdot)$ is a bivariate spline function. We used the first two dimensions of the GPSs to construct the spline function in our simulations and data analysis. An empirical sensitivity analysis found the estimates of causal effects did not vary with the dimensions of the GPSs; see Table S5 in the Supplemental Materials. The multivariate spline function can be specified using additive spline bases, tensor products of spline bases or radial basis functions \citep{ruppert2003semiparametric}. To better capture the correlation between the components of $\bm{R}(\bm{X}_i)$ and facilitate easy-implementation of this method, we consider a tensor product spline that can be implemented using the off-the-shelf statistical software packages for $h(\cdot)$. The estimator of the ATE $\Delta_{k,l}$ by this method is 
\begin{equation}
\widehat{\Delta}^{rams}_{k,l} = \frac{\sum_{i=1}^N f(k,\bm{X}_i)}{\sum_{i=1}^N f(l,\bm{X}_i)} = \frac{\sum_{i=1}^N \text{logit}^{-1} \bigg{\{}  \hat{\beta}_k I(W_i=k) + h(\bm{R}(\bm{X}_i),\hat{\bm{\gamma}}) \bigg{\}}}{\sum_{i=1}^N \text{logit}^{-1} \bigg{\{}  \hat{\beta}_l I(W_i=l) + h(\bm{R}(\bm{X}_i),\hat{\bm{\gamma}}) \bigg{\}}}.
\end{equation} 
In our simulations and data application, we also considered a variant of RAMS with trimmed GPSs to lessen the impact of extreme GPSs.  

\section{Simulation studies}

\subsection{Simulation design}
We conducted three sets of simulations to investigate and compare the operating characteristics of the methods described in Section 2.2. We based the design of our simulations on the structure of the SEER-Medicare data described in Section 1.3. We assumed three treatment groups ($Z=3$) throughout the simulations.

\subsubsection{Simulation I: Ratio of units and sample size}
In the first simulation, we considered a combination of two design factors: the ratio of units in the treatment groups ($n_1:n_2:n_3$) and the study sample size (i.e., the total number of units $N = n_1+n_2+n_3$). We considered three scenarios for the two factors: 1) $n_1:n_2:n_3 =1:1:1$ and $N = 1500$, 2) $n_1:n_2:n_3 =1:4:3$ and $N=4000$ and 3) $n_1:n_2:n_3 =1:10:8$ and $N = 9500$ to respectively represent equal, moderately unequal and highly unequal sample sizes across treatment groups. The third scenario of highly unequal sample sizes represents the SEER-Medicare database study, in which RAS is the smallest treatment group for it is a relatively new technology. 

We considered a total of 10 confounders consisted of five continuous variables and five categorical variables.  We assumed that both the treatment assignment mechanism and the response surfaces are nonlinear models of the confounders. Specifically, the treatment assignment followed a multinomial logistic regression model. Assuming  the third treatment group is the reference group, $W_i \sim \text{Multinomial}(N,p_1,p_2,p_3)$ with 
\begin{equation}
\label{eq:trtmod}
\begin{split}
p(W_i=1|\bm{X}_i) &= \frac{ex_{1i}}{1 + ex_{1i} + ex_{2i}}\\
p(W_i=2|\bm{X}_i) &= \frac{ex_{2i}}{1 + ex_{1i} + ex_{2i}}\\
p(W_i=3|\bm{X}_i) &= \frac{1}{1 + ex_{1i} + ex_{2i}},
\end{split}
\end{equation}
and 
\begin{equation*}
\begin{split}
ex_{1i} &= \exp(\alpha_1 + \bm{X}_i\xi_1^L + \bm{Q}_i\xi_1^{NL})\\
ex_{2i} &= \exp(\alpha_2 + \bm{X}_i\xi_2^L + \bm{Q}_i\xi_2^{NL}),
\end{split}
\end{equation*}
where $\bm{Q}_i$ denotes the nonlinear transformations and higher-order terms of the covariates in $\bm{X}_i$, $\xi_1^L$ and $\xi_2^L$ are vectors of coefficients associated with $\bm{X}_i$, and $\xi_1^{NL}$ and $\xi_2^{NL}$ are vectors of coefficients associated with $\bm{Q}_i$. The intercepts $\alpha_1$ and $\alpha_2$ were specified to create the corresponding ratio of units in three treatment groups in each scenario. We generated three sets of response surfaces as follows:
\begin{equation}
\label{eq:po}
E[Y(w) | \bm{X}_i] = \text{logit}^{-1}  \{\tau_w + \bm{X}_i\eta_w^{L} + \bm{Q}_i \eta_w^{NL} \},  \quad  \; w \in \{1,2,3\},
\end{equation}
where regression coefficients ($\tau_w$, $\eta_w^L$, and $\eta_w^{NL}$) were chosen so that the outcome prevalence in the treatment groups were similar as the rates of cardiovascular complication (2\%--3\%) observed in the SEER-Medicare data. Details of model specification in Equation~\eqref{eq:trtmod} and~\eqref{eq:po} appear in Table S1 of Supplementary Materials.

We compared the following nine methods in Simulation I: 1) IPTW with GPS estimated using multinomial logistic regression (IPTW-MLR); 2) IPTW with GPS estimated using generalized boosted models (IPTW-GBM); 3) IPTW-MLR with trimmed weights; 4) IPTW-GBM with trimmed weights; 5) RAMS with GPS estimated using multinomial logistic regression (RAMS-MLR); 6) RAMS with GPSs estimated using the generalized boosted models (RAMS-GBM); 7) RAMS-MLR with trimmed GPSs (RAMS-MLR-Trim); 8) RAMS-GBM with trimmed GPSs (RAMS-GBM-Trim); and 9) BART.

\subsubsection{Simulation II: Outcome prevalence}
We further compared the nine methods vetted in the first simulation with respect to another design factor: the outcome prevalence. We varied the outcome prevalence in each of three treatment groups in two scenarios: 1) $1\% - 5\%$ (extremely rare), and 2) $5\% - 10\%$ (rare). That is, we simulated the outcomes with the prevalence rate falling within the range 1-5\%  and 5-10\%, respectively. We generated data sets following the simulation configuration of scenario 3 in Simulation I, that is, $N = 9,500$ and $n_1:n_2:n_3 = 1:10:8$. The model specification of the treatment assignment mechanism is the same as in Simulation I and $\tau_w$'s in the response surfaces were modified to vary the outcome prevalence.

\subsubsection{Simulation III: Covariate overlap}
We finally investigated how levels of covariate overlap impact causal estimates with multiple treatments and rare outcomes. We generated data sets using the simulation configuration of scenario 3 in Simulation I and of scenario 1 in Simulation II, including the total sample size, the ratio of units, the outcome prevalence, the number of continuous and categorical confounders, and the response surface models, to mimic the SEER-Medicare dataset, but varied the treatment assignment models~\eqref{eq:trtmod} to create shifting degrees of covariate overlap.

We considered three levels of covariate overlap: 1) \emph{strong} - there was strong overlap with respect to each of the 10 confounders; 2) \emph{moderate} - the five categorical variables had sufficient overlap as in the \emph{strong} scenario but there were lack of overlap in the five continuous variables ; and 3) \emph{weak} - there was lack of overlap in the covariate space defined by all 10 confounders. 
In all scenarios, the lack of overlap was induced in true confounders, i.e., variables included in both the treatment assignment model and the response surfaces.  This simulation was designed to make it difficult for any method to successfully estimate the true treatment effects, as both the treatment assignment and the outcome are difficult to model. 

We first generated the treatment variable $W$ from a multinomial distribution, $W \sim \text{Multinomial } (9500,\\ .05, .53, .42)$, and then generated each  covariate from their respective distributions conditional on treatment assignment to create varying levels of covariate overlap. In the \emph{strong} scenario overlap, we created strong covariate overlap by generating similar distributions of the covariates across the treatment groups for all 10 confounders. Specifically, we assumed that the five continuous variables follow the normal distribution $X_{ij} | W_i \sim \mathcal{N}(0,1-0.01W_i)$ for $j = 1, \ldots, 5$, and the five categorical variables follow multinomial distribution $X_{ij} | W_i \sim \text{Multinomial} (N, 0.3, 0.3,  0.4)$ for $j = 6, \ldots, 10$. 
In the \emph{moderate} overlap scenario, we assumed that the five continuous variables follow the normal distribution $X_{ij} | W_i \sim \mathcal{N}(0.05W_i,1-0.05W_i)$ for $j = 1, \ldots, 5$, and the five categorical variables follow the multinomial distribution $X_{ij} | W_i \sim \text{Multinomial} (N, 0.3 - 0.01W_i, 0.3 +0.01W_i, 0.4)$ for $j = 6, \ldots, 10$. 
 In the \emph{weak} scenario overlap,  we created the lack of overlap for five continuous variables. We assumed that $X_{ij} | W_i = 1\sim \mathcal{N}(-0.5,1)$, $X_{ij} | W_i = 2\sim \mathcal{N}(1,1)$, $X_{ij} | W_i = 3\sim \mathcal{N}(2,1)$ for $j =1, \ldots, 5$, and $X_{ij} | W_i \sim \text{Multinomial} (N, 0.3-0.001W_i, 0.3+0.001W_i, 0.4)$, for $j = 6, \ldots, 10$.
 Boxplots of distributions of estimated GPSs across the treatments were used to visualize the degree of covariate overlap, see Figure S1 in Supplemental Materials. We used the BART discarding strategy in~\eqref{eq:disc} to assess whether this strategy (BART-Discard) improved over BART.

For all three simulations described above, we evaluated the performance of all methods considered using the ATE estimates based on marginal RD.  True treatment effects were computed based on a simulated population of size 100,000. For each method, in each of 200 replications of each scenario, we estimated three pairwise ATEs, $ATE_{1,2}$, $ATE_{1,3}$ and $ATE_{2,3}$, and the bias in each estimate. Using these values, we calculated for each method and each scenario, the mean absolute bias (MAB) and the root mean square error (RMSE). The Monte Carlo simulation error (MCSE) was also provided to show the variability across simulation replications. The number of events and outcome prevalence rates in each of treatment groups for each of simulation scenarios are summarized in Table S4 in Supplemental Materials.

We implemented the methods as follows. The GPSs were estimated using both multinomial logistic regression model (MLR) and GBM. GBM was implemented using the \texttt{mnps()} function available in the \texttt{twang} package in \verb+R+. The The stopping rule for the optimal iteration of the generalized boosted model was based on maximum of absolute standardized bias, which compares the distributions of the covariates between treatment groups \citep{mccaffrey2013tutorial}. RAMS was implemented using the \texttt{gam()} function with tensor product smoother \texttt{te()} in the \texttt{gam} package in R. The weights or the GPSs were trimmed at 5\% and 95\% to generate trimmed estimators for IPTW and RAMS, respectively. For BART, we used the default priors associated with the \texttt{bart()} function available in the \texttt{BART} package in \verb+R+. For each BART fit, we allowed the maximum number of trees in the sum to be 100. To ensure the convergence of the MCMC in BART, we let the algorithm run for 5000 iterations with the first 3000 considered as burn-in.

\subsection{Simulation Results}

\subsubsection{Simulation I}
Table~\ref{simulation1} displays the MAB and RMSE of the estimates of 
three ATE effects $ATE_{1,2}$,$ATE_{1,3}$ and $ATE_{2,3}$ for the three scenarios in Simulation I. BART consistently yielded the lowest MAB and RMSE across all three scenarios with different ratio of units and varying sample sizes, and for all pairwise treatment effects. The second best method is RAMS, judged by both MAB and RMSE. Implementation of the GBM for estimating the GPSs only moderately improved the performance of RAMS over the use of MLR. The flexible spline function of the GPSs in RAMS may  lessen the need for GPS model accuracy. 
IPTW based methods produced large MAB and RMSE, with IPTW-MLR giving the largest bias and variability.  Trimming appeared to improve the performance of IPTW-MLR but had negligible effect on IPTW-GBM and RAMS. BART was not sensitive to the ratio of units or sample size, whereas the GPS methods tended to have better performance as the sample sizes in the comparison treatment groups grew relative to the reference treatment group. 

\begin{table}[H]
\centering
\scriptsize
\caption{Comparisons of the estimated average treatment effects (ATE) in terms of mean absolute bias (MAB) and root mean square error (RMSE) across 200 replications in Simulation I. The causal estimand is based on percent risk difference, i.e.,  (risk difference between treatment groups)$\times 100$. MSCE: Monte Carlo simulation error, calculated as standard deviation of bias/$\sqrt{200}$. In Scenario 1), the true $ATE_{1,2} =0.88\%$, $ATE_{1,3} = 0.70\%$, $ATE_{2,3} =0.92\%$. In Scenario 2), the true $ATE_{1,2} =0.86\%$, $ATE_{1,3} = 0.74\%$, $ATE_{2,3} =0.98\%$. In Scenario 3), the true $ATE_{1,2} =0.82\%$,$ATE_{1,3} = 0.74\%$, $ATE_{2,3} =0.94\%$.}
\begin{tabular}{clcccccccccccc}
\hline
& &  \multicolumn{3}{c}{$\text{ATE}_{1,2}$} && \multicolumn{3}{c}{$\text{ATE}_{1,3}$}&& \multicolumn{3}{c}{$\text{ATE}_{2,3}$}\\
\cline{3-5} \cline{7-9}\cline{11-13}
Scenario & Method & MAB & RMSE & MCSE && MAB & RMSE & MCSE && MAB & RMSE & MCSE \\
\hline
& IPTW-MLR & 3.10 & 3.44 & 0.24 && 2.91 & 3.31 & 0.23 && 2.92 & 3.31 & 0.23\\  & IPTW-MLR-Trim & 2.46 & 2.81 & 0.20 && 2.53 & 2.80 & 0.20 && 2.51 & 2.80 & 0.20\\  & IPTW-GBM & 2.14 & 2.47 & 0.17 && 1.93 & 2.21 & 0.16 && 2.09 & 2.21 & 0.16\\  & IPTW-GBM-Trim & 1.95 & 2.22 & 0.16 && 1.92 & 2.18 & 0.15 && 1.86 & 2.18 & 0.15\\ 1) $n=1500$ & BART & 1.02 & 1.15 & 0.08 && 0.80 & 0.92 & 0.06 && 0.88 & 0.92 & 0.06\\ Ratio of units = 1:1:1 & RAMS-MLR & 1.62 & 1.81 & 0.12 && 1.44 & 1.61 & 0.11 && 1.45 & 1.61 & 0.11\\  & RAMS-MLR-Trim & 1.52 & 1.69 & 0.12 && 1.35 & 1.52 & 0.11 && 1.42 & 1.52 & 0.11\\  & RAMS-GBM & 1.57 & 1.76 & 0.12 && 1.33 & 1.51 & 0.11 && 1.46 & 1.51 & 0.11\\  & RAMS-GBM-Trim & 1.46 & 1.72 & 0.12 && 1.25 & 1.43 & 0.10 && 1.40 & 1.43 & 0.10  \\
\hline
 & IPTW-MLR & 2.44 & 2.74 & 0.19 && 2.49 & 2.80 & 0.20 && 2.63 & 2.80 & 0.20\\  & IPTW-MLR-Trim & 2.07 & 2.32 & 0.16 && 1.95 & 2.19 & 0.15 && 2.02 & 2.19 & 0.15\\  & IPTW-GBM & 1.50 & 1.70 & 0.12 && 1.50 & 1.70 & 0.12 && 1.53 & 1.70 & 0.12\\  & IPTW-GBM-Trim & 1.57 & 1.77 & 0.12 && 1.40 & 1.58 & 0.11 && 1.29 & 1.58 & 0.11\\ 2) $n=4000$ & BART & 0.89 & 0.99 & 0.07 && 0.89 & 1.01 & 0.07 && 1.00 & 1.01 & 0.07\\ Ratio of units = 1:4:3 & RAMS-MLR & 1.25 & 1.43 & 0.10 && 1.22 & 1.37 & 0.10 && 1.17 & 1.37 & 0.10\\  & RAMS-MLR-Trim & 1.17 & 1.33 & 0.09 && 1.29 & 1.45 & 0.10 && 1.15 & 1.45 & 0.10\\  & RAMS-GBM & 1.35 & 1.50 & 0.11 && 1.16 & 1.31 & 0.09 && 1.20 & 1.31 & 0.09\\  & RAMS-GBM-Trim & 1.18 & 1.35 & 0.09 && 1.19 & 1.34 & 0.09 && 1.33 & 1.34 & 0.09  \\
\hline
& IPTW-MLR & 2.04 & 2.27 & 0.16 && 1.99 & 2.17 & 0.15 && 1.98 & 2.17 & 0.15\\  & IPTW-MLR-Trim & 1.75 & 1.98 & 0.14 && 1.95 & 2.14 & 0.15 && 1.77 & 2.14 & 0.15\\  & IPTW-GBM & 1.35 & 1.51 & 0.10 && 1.36 & 1.53 & 0.11 && 1.36 & 1.53 & 0.11\\  & IPTW-GBM-Trim & 1.25 & 1.41 & 0.10 && 1.21 & 1.35 & 0.09 && 1.24 & 1.35 & 0.09\\ 3) $n=9500$ & BART & 0.90 & 1.03 & 0.07 && 0.82 & 0.90 & 0.06 && 0.74 & 0.90 & 0.06\\ Ratio of units = 1:10:8  & RAMS-MLR & 1.14 & 1.26 & 0.09 && 1.05 & 1.20 & 0.08 && 1.04 & 1.20 & 0.08\\  & RAMS-MLR-Trim & 1.11 & 1.27 & 0.09 && 1.02 & 1.17 & 0.08 && 0.95 & 1.17 & 0.08\\  & RAMS-GBM & 1.09 & 1.23 & 0.09 && 0.94 & 1.08 & 0.07 && 1.00 & 1.08 & 0.07\\  & RAMS-GBM-Trim & 1.09 & 1.23 & 0.09 && 0.95 & 1.09 & 0.08 && 1.03 & 1.09 & 0.08 \\
\hline
\end{tabular}
\label{simulation1}
\end{table}

\subsubsection{Simulation II}
\begin{figure}
    \centering
    \includegraphics[scale=0.15]{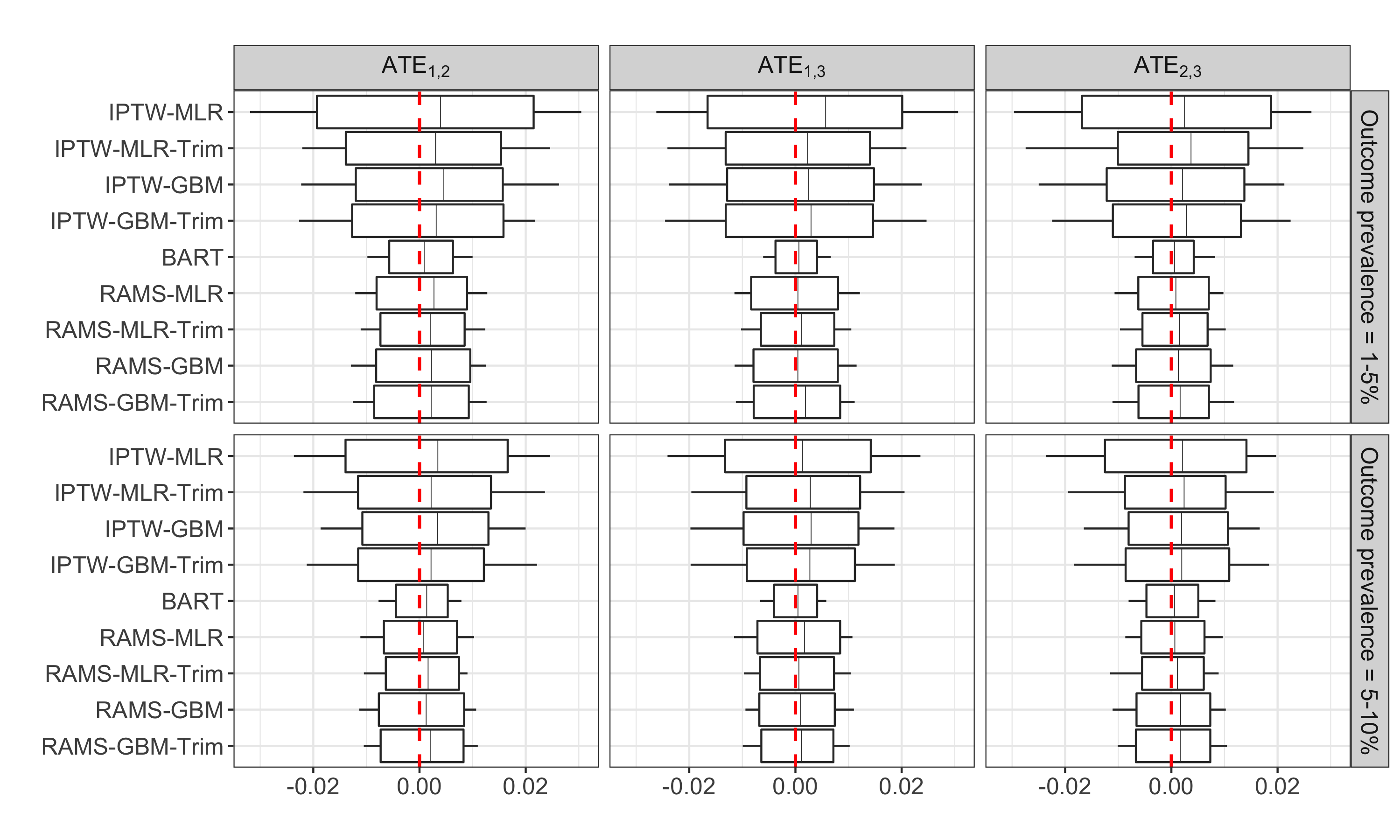}
    \caption{Biases among 200 replications under two outcome prevalence rates of 1--5\% and 5--10\% for each of nine methods and three ATE effects ${ATE}_{1,2}$, ${ATE}_{1,3}$ and ${ATE}_{2,3}$ based on percent risk difference, i.e.,  (risk difference between treatment groups)$\times 100$. For  prevalence rate of 1--5\%, the true $ATE_{1,2} = 0.73\%$, $ATE_{1,3} = 1.17\%$,  $ATE_{2,3 } = 1.21\%$. For  prevalence rate of 5--10\%, the true $ATE_{1,2} = 1.26\%$, $ATE_{1,3} = 0.72\%$,  $ATE_{2,3 } = 0.98\%$.  }
    \label{fig:sim2}
\end{figure}

Figure~\ref{fig:sim2} displays boxplots of biases based on RD of three treatment effects  $ATE_{1,2}$,  $ATE_{1,3}$ and $ATE_{2,3}$ among 200 simulations for each of the nine methods and two outcome prevalence rates of 1\%--5\% and 5\%--10\% in Simulation II. As in Simulation I, BART boasted the smallest bias and variability among the nine methods for both outcome prevalence rates across all three treatment effects. RAMS, following BART, delivered considerably better performance than IPTW based methods, for having both smaller bias and higher precision. Using GBM to estimate the GPSs for RAMS led to only a moderate bias reduction compared to using MLR, with slightly better improvement when the outcome prevalence was lower. Trimming again improved only IPTW-MLR but did not have a noticeable impact on GBM and RAMS. The advantage of BART over RAMS diminished when the outcome prevalence increased and RAMS achieved better bias reduction and precision. IPTW based methods also had better performance under higher outcome prevalence.  Table S2 in Supplemental Materials compares the  MAB and RMSE of each method under two outcome prevalence rates. An alternative representation of the biases based on the relative risk (RR) appears in Figure S3 in Supplemental Materials. The RR biases show similar comparative performance of the nine methods as the RD biases. The density distributions of RR biases appear to be approximately normal, whereas the density distributions of RD biases are bimodal with the curves flattening around zero, as demonstrated in Figure S4 of the Supplemental Materials. The difference in distributional shapes could be due to rare outcomes and very small true treatment effects.

\subsubsection{Simulation III}
\begin{figure}
    \centering
    \includegraphics[scale=0.15]{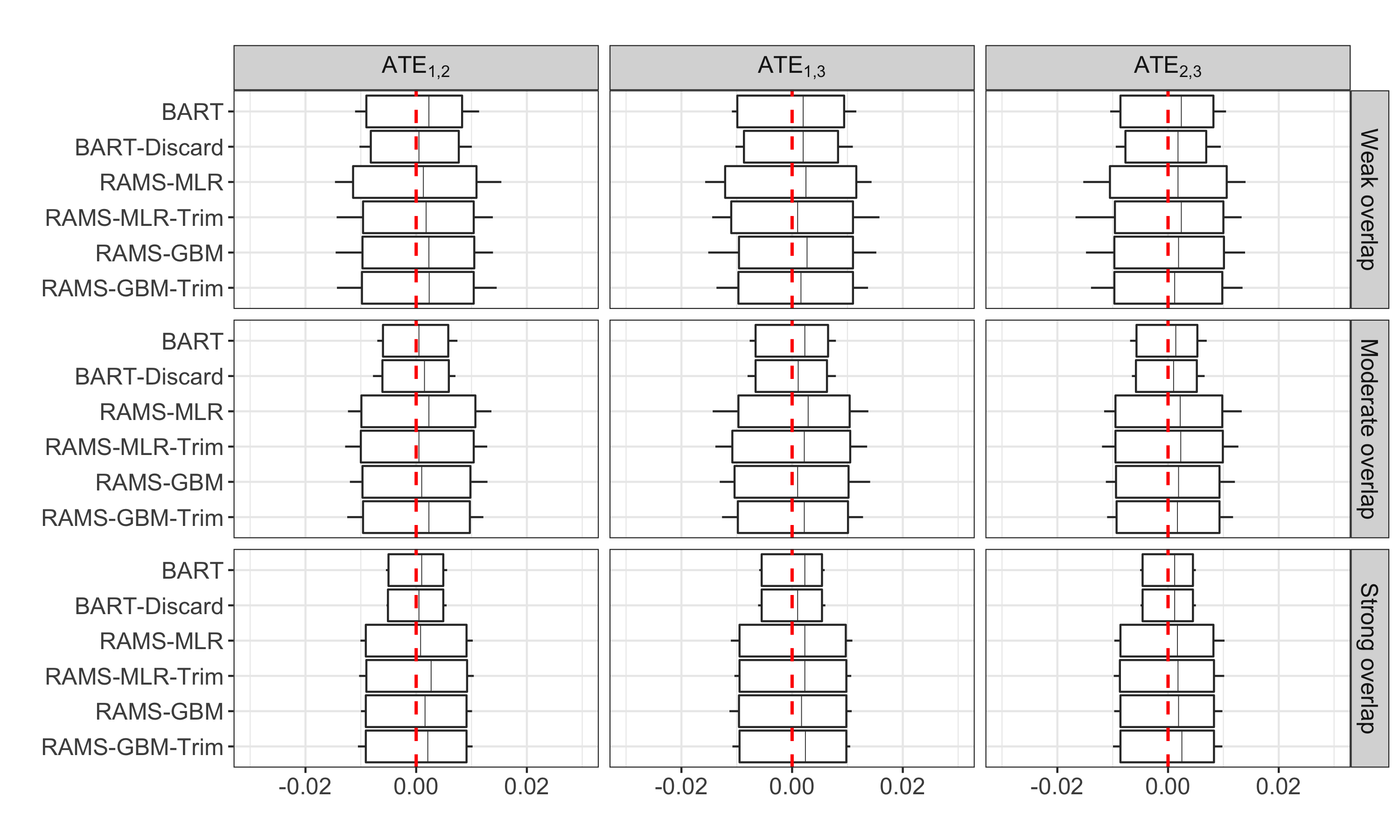}
    \caption{Biases among 200 replications under varying level of covariate overlap for BART vs. RAMS and three treatment effects $ATE_{1,2}$, $ATE_{1,3}$ and $ATE_{2,3}$. The causal estimand is the marginal percent risk difference between treatment groups, i.e., (risk difference between treatment groups) $\times 100$. For weak overlap, the true $ATE_{1,2} = 1.25\%$, $ATE_{1,3} = 0.77\%$,  $ATE_{2,3 } = 0.88\%$. For moderate overlap, the true $ATE_{1,2} = 0.97\%$, $ATE_{1,3} = 0.89\%$,  $ATE_{2,3 } = 1.16\%$. For strong overlap, the true $ATE_{1,2} = 0.97\%$, $ATE_{1,3} = 0.73\%$,  $ATE_{2,3 } = 1.40\%$. }
    \label{fig:sim3}
\end{figure}

Figure~\ref{fig:sim3} compares biases of three treatment effect estimators ${ATE}_{1,2}$,  ${ATE}_{1,3}$ and ${ATE}_{2,3}$ among 200 simulations under three levels of covariate overlap for each of RAMS-MLR, RAMS-MLR-Trim, RAMS-GBM, RAMS-GBM-Trim, BART and BART with discarding rules. 

BART demonstrated smaller bias and variability than RAMS across all three levels of covariate overlap. The difference is less evident in the scenario of weak covariate overlap than in the scenarios of moderate and strong overlap. The BART discarding rule  considerably reduced the biases in the estimates of  the ATE effects in the weak scenario where there was substantial lack of covariate overlap. When there was a strong or moderate covariate overlap, BART with and without discarding performed equally well. BART discarding strategy excluded  7.1\% , 0.2\% and 0\%  units on average across 200 replications for the weak, moderate and strong overlap, respectively. Flexibility in estimating the GPSs using the GBM  improved the performance of RAMS when the covaraite overlap was weak or moderate. Trimming improved RAMS-MLR under weak overlap, and gave similar results when the covariate overlap was strong or moderate. Table S3 in Supplemental Materials shows the MAB and RMSE of each method under different levels of covariate overlap.  We also plotted the RR biases in Figure S5 and the density distributions of RR and RD biases in Figure S6 of the Supplemental Materials. The boxplots of RR biases demonstrate similar comparative performance of the six methods as the RD biases. The difference between distributional shapes of the two types of biases is the same as demonstrated in Simulation II.

\section{Real data analysis}\label{sec:application}

As mentioned in Section 1.3, in this paper we compared the effectiveness of the three surgical procedures, RAS, VATS and OT, on the following four postoperative outcomes: the presence of (i) extra-pulmonary infection; (ii) cardiovascular complication; (iii) thromboembolic complication; and (iv) reoperation within 30 days of surgery or during the hospitalization in which the primary surgical procedure was performed. 

We estimated three pairwise ATEs in terms of marginal RD between the three surgical procedures in the overall population using the methods examined in the simulations. IPTW-MLR was not included in the real data application due to its relatively poor performance shown in the simulations. In addition, trimming method was not considered for RAMS because the impact of trimming on the GPS for RAMS was shown to be trivial. Results from BART-Discard were not shown as they were the same as the estimates of BART (no units were discarded). An assessment of covariate overlap shows that the baseline characteristics has relatively strong overlap; see Figure S2 in Supplemental Materials. Confidence intervals of ATEs were estimated using nonparametric bootstrap for IPTW and RAMS, and Bayesian posterior credible intervals were estimated for BART. All of the methods were implemented as described in the simulation studies.

Table~\ref{Tab.Seer.Res} presents the point and interval estimates of the ATEs for RAS versus OT, RAS versus VATS and OT versus VATS, in terms of marginal RD. All of the methods yielded statistically insignificant treatment effects on the four outcomes for RAS versus OT or VATS, suggesting that RAS did not have a significant advantage over VATS or OT on postoperative complications. Results from different methods consistently showed lower rates of post-operative extrapulmonary infection and reoperation among patients treated with VATS compared to OT. Of the four methods, BART provided the narrowest uncertainty intervals, whereas IPTW produced the largest width of confidence intervals. Results from this comparative effectiveness study provided partial evidence that VATS may lead to lower rates of post-operative reoperation and extrapulmonary infection compared to OT, and that RAS has no evident advantage over VATS or OT. 

\begin{table}[htbp]
\centering
\footnotesize
\caption{\footnotesize{Estimated $ATE_{1,2}$,  $ATE_{1,3}$ and $ATE_{2,3}$ (95$\%$ uncertainty intervals in parentheses) of three surgical procedures on postoperative rare outcomes. The causal estimand is the marginal percent risk difference, i.e.,  (risk difference between treatment groups) * 100.}}\label{Tab.Seer.Res}
\begin{tabular}{clcccccc}
\cline{1-8} 
&& \multicolumn{2}{c}{RAS vs. OT} & \multicolumn{2}{c}{RAS vs. VATS} &
\multicolumn{2}{c}{OT vs. VATS} \\
\cline{3-8} 
 Outcomes & Methods & Estimate & 95$\%$ CI & Estimate & 95$\%$ CI & Estimate & 95$\%$ CI\\

\hline & IPTW-GBM & -2.9 & (-5.5,3.0) & -0.6 & (-3.2,2.3) & 2.3 & (0.7,3.8)\\ Extrapulmonary & BART & -2.1 & (-3.2,0.4) & -0.4 & (-1.8,1.0) & 1.7 & (1.0,2.4)\\ infection & RAMS-MLR & -2.7 & (-4.3,0.6) & -0.7 & (-2.3,1.3) & 1.9 & (0.7,3.5)\\  & RAMS-GBM & -2.9 & (-4.5,0.9) & -0.7 & (-2.2,1.3) & 2.3 & (1.6,3.6)\\  
\hline  & IPTW-GBM & 0.1 & (-1.9,5.7) & 0.6 & (-1.3,6.6) & 0.5 & (-0.8,1.8)\\ Cardiovascular & BART & 0.1 & (-1.0,1.4) & 0.2 & (-0.5,1.0) & 0.2 & (-0.1,0.6)\\ complication & RAMS-MLR & 0.5 & (-1.1,2.2) & 0.8 & (-0.8,2.4) & 0.3 & (-0.2,1.0)\\  & RAMS-GBM & 0.1 & (-1.3,1.6) & 0.5 & (-0.9,2.0) & 0.4 & (-0.1,1.0)\\  
\hline & IPTW-GBM & 0.7 & (-2.5,6.4) & 0.6 & (-2.8,5.9) & -0.1 & (-2.1,1.5)\\ Thromboembolic & BART & 0.6 & (-1.3,2.6) & 0.5 & (-1.2,2.3) & 0.0 & (-0.5,0.5)\\ complication & RAMS-MLR & 0.8 & (-1.7,3.3) & 0.9 & (-1.5,3.4) & 0.1 & (-0.7,0.9)\\  & RAMS-GBM & 0.7 & (-1.5,3.1) & 0.6 & (-1.7,3.0) & -0.1 & (-1.0,0.7)\\  
\hline & IPTW-GBM & -0.3 & (-1.6,3.2) & 0.4 & (-0.9,3.4) & 0.7 & (0.1,1.3)\\ Reoperation & BART & -0.3 & (-1.1,0.9) & 0.3 & (-0.3,0.9) & 0.5 & (0.3,0.7)\\  & RAMS-MLR & -0.3 & (-1.3,1.0) & 0.5 & (-0.5,1.7) & 0.8 & (0.4,1.2)\\  & RAMS-GBM & -0.3 & (-1.2,1.0) & 0.5 & (-0.4,1.7) & 0.8 & (0.4,1.1)\\
  \hline
\end{tabular}
\end{table}

We performed a sensitivity analysis to evaluate the impact of the choice of hyperparameter in the end-node piror for BART, that is, the choice of $k$ in the \texttt{bart()} function \citep{dorie2016flexible}. Five-fold cross-validation was used to select the optimal hyperparameter $k$ that minimized the misclassification error. Results suggest that BART with the optional $k$ gave similar point and interval estimates of the ATEs as BART with the default value of $k$ (not shown). Furthermore, we implemented our BART-specific discarding rule for ATEs \eqref{eq:disc}  to identify if there are any units that lack common support and thus should be discarded to avoid inference for them. We did not exclude any such patients from the real data set.   We conducted an empirical investigation to examine whether the dimensions of the GPSs for the bivariate spline function of the RAMS method impact the estimation of the causal effects. Table S5 shows that the estimated ATEs did not change with different dimensions of the GPSs.

\section{Discussion}
The primary contribution of this paper to the causal inference literature is the proposal and extension of several methods to the multiple treatments and rare outcome setting. The advent of large-scale population-based healthcare database brings analytic challenges and demands novel methodology beyond traditional causal inference techniques. Causal inference methods have been predominantly focusing on a binary treatment, and research for handling multiple treatments especially when the outcome is a rare event is sparse. Previous studies have provided extensive empirical evidence for the comparison of a variety of causal inference methods in either the multiple treatment setting \citep{hu2020estimation} or the rare outcome setting \citep{franklin2017comparing}, but not both. In the multiple treatments setting, \cite{hu2020estimation} explicated that BART gives lower bias, smaller RMSE, more coherent uncertainty intervals and better convergence property than other methods examined. \cite{franklin2017comparing} demonstrated that regression on the PS using a spline function produces smaller bias and lower standard errors than other PS-based methods in the context of rare outcomes.

We extended both BART and regression adjustment with PS using spline to the multiple treatment setting when the outcome is binary with low prevalence, and compared them with the commonly used IPTW-based methods. We conducted three unique sets of simulations that represented complex causal inference settings motivated by the structure of the SEER-Medicare data. 

BART consistently produced better performance in terms of bias and RMSE for pairwise ATEs across all scenarios regarding ratio of units, outcome prevalence and covariate overlap. RAMS, using a multidimensional spline of GPS, trailed BART but outperformed all IPTW-based methods including the machine learning based IPTW-GBM. The performance of RAMS was generally improved when more flexible models were used to estimate the GPS. The advantage of BART over RAMS was reduced and both methods delivered good performance when the outcome prevalence was relatively higher (5\%--10\%) or when there was sufficient covariate overlap.  When there is a substantial amount of lack of overlap in the covariate space, the BART-specific discarding rule markedly improved over plain BART. 

RAMS-MLR is the most computationally efficient method while IPTW-GBM is the least. On a dataset of size $n=9500$, one RAMS-MLR implemention takes less than 5 seconds to run and one BART implementation takes less than 150 seconds on a iMAC with a 4GHz Intel Core i7 processor. In a stark comparison, one IPTW-GBM implementation takes about 10 minutes to generate results. Considering the easy implementation and computational efficiency, RAMS may be recommended in the settings where the covariate overlap is strong or the outcome prevalence is relatively high.

We implemented methods examined on 11980 stage I-IIIA NSCLC patients, drawn from the latest SEER-Medicare linkage. Results suggest that VATS may be preferred over OT in terms of extrapulmonary infection and reoperation in the general population. There is no evidence showing that RAS leads to better post-operative outcomes compared to OT or VATS. Given that RAS was
associated with significantly higher costs, specifically those incurred in the pre-operative period compared to VATS \citep{veluswamy2019comparative}, RAS therefore may not be a cost-effective option in routine care for the surgical management of patients with NSCLC.

Our work provides several important avenues for future research in the causal inference settings of multiple treatments and rare outcomes.  First, the flexibility offered by the nonparametric modeling of BART can be leveraged to model nonlinear regression relationship in time to event data. Second, individual treatment effects and coherent uncertainty intervals can be easily obtained from the BART model, which provides a building block for assessing heterogeneous treatment effect and identify subgroups for optimal treatment. Third, the proposed RAMS assumes that the treatment variable and spline function of the GPS are included into the model additively. This model formulation may not provide a valid treatment effect estimate in the presence of treatment effect heterogeneity because it ignores potential interaction between the treatment variable and the spline function of the GPS. \citet{gutman2013robust} proposed a multiple imputation with two splines and subclassification method (MITSS) for a binary treatment, which can correctly model the association between the outcome and the PS even in the presence of strong treatment heterogeneity. Further development of RAMS for estimating heterogeneous treatment effects is a future direction. Finally, we have made a significant untestable assumption related to unmeasured confounding. Developing sensitivity analyses under the complex multiple treatment and rare outcome setting leveraging BART would also be a worthwhile and important contribution \citep{hogan2014bayesian,hu2017modeling}.

\begin{acknowledgements}
The authors are grateful to Jiayi Ji for the help with replicating simulations and organizing the results, and to the editor, associate editor, and two anonymous reviewers whose comments and suggestions led to substantial improvements in the manuscript. This study used the linked SEER-Medicare database. The interpretation and reporting of these data are the sole responsibility of the authors. The authors acknowledge the efforts of the National Cancer Institute; the Office of Research, Development and Information, CMS; Information Management Services (IMS), Inc.; and the Surveillance, Epidemiology, and End Results (SEER) Program tumor registries in the creation of the SEER-Medicare database.
\end{acknowledgements}

\noindent \small \textbf{Funding} This work was in part supported by award ME\_2017C3\_9041 from the Patient-Centered Outcomes Research Institute, and by grants R21CA245855 and P30CA196521-01 from the National Cancer Institute.\\

\noindent \small \textbf{Conflict of interest} The authors  declare that they have no conflict of interest. \\

\noindent \small \textbf{Ethical approval} This article involves  only simulated data and secondary analysis of deidentified data, and does not involve intervention or interaction with living individuals, or identifiable private information. The institutional review board at the Icahn School of Medicine with which the lead author is affiliated has waived the need for informed consent.

%
%

\bibliographystyle{spbasic}      
\bibliography{MultiTrts_rare_manuscript}   

\end{document}